\documentclass[structabstract]{aa}  
\usepackage{graphicx}
\usepackage{amsmath}
\usepackage{amssymb}
\usepackage{txfonts}               
\usepackage{longtable,lscape}
\usepackage{hyperref}
\usepackage{natbib}                
\newcommand{\be}{\begin{eqnarray}}
\newcommand{\ee}{\end{eqnarray}}
\newcommand{\beq}{\begin{equation}}
\newcommand{\eeq}{\end{equation}}
\begin{document}
   \title{A two-component model for the high-energy variability of blazars. Application to PKS 2155-304}


\author{
  M.~M.~Reynoso\inst{1}
  \and
  G.~E.~Romero\inst{2}
  \and
  M.~C.~Medina\inst{2,3}
}

\institute{Instituto de Investigaciones F\'{\i}sicas de Mar del Plata (CONICET - UNMdP), Facultad de Ciencias Exactas y Naturales, Universidad Nacional de Mar del Plata, Dean Funes 3350, (7600) Mar del Plata, Argentina \and Instituto Argentino de Radioastronom\'ia, CCT La Plata-CONICET, 1894, Villa Elisa C.C. No. 5, Argentina \and Irfu, Service de Physique des Particules, CEA Saclay,F-91191 Gif-sur-Yvette Cedex, France}


 
  \abstract
   {}
   {We study the production of VHE emission in blazars as a superposition of a steady 
component from a baryonic jet and a time-dependent contribution from an inner $e^-e^+$ beam
launched by the black hole.}
  {Both primary relativistic electrons and protons are injected in the jet, and the particle distributions along it are found by solving a one-dimensional transport equation that accounts for convection and cooling. The short-timescale variability of the emission is explained by local pair injections in turbulent regions of the inner beam.}
   {For illustration, we apply the model to the case of PKS 2155-304, reproducing a quiescent state of emission with inverse Compton and synchrotron radiation from primary electrons, as well as proton-proton interactions in the jet. The latter also yield an accompanying neutrino flux that could be observed with a new generation km-scale detector in the northern hemisphere such as KM3NeT.}
   {}


   \maketitle
%

\section{Introduction}

Blazars are the AGNs in which the jet points mainly in the direction of
the line of sight. They exhibit the most extreme high-energy phenomena of all AGNs.
Their spectral energy distributions (SEDs) are characterized by nonthermal continuum spectra with a broad low-frequency component from X-rays to $\gamma$-rays. Blazars show rapid variability across the entire electromagnetic spectrum. Variability at high energies on timescales of a few minutes has been observed for some of them, such as PKS 2155-304 (e.g. Aharonian et al. 2006).

In this work we present a two-{component} jet model with both relativistic leptons and hadrons to explain the high-energy emission
from these objects. The basic scenario consists of a steady baryonic jet launched by the accretion disk, and an $e^+e^-$ beam launched by the black hole ergosphere. The quiescent component of the signal is assumed to 
be produced by the jet, while the variable component is due to shocked regions in the inner $e^+e^-$ beam. Inhomogeneities and turbulence can be generated by Kelvin-Hemholtz instabilities. In Sect. 2 we describe the basics of the model. Its application to PKS 2155-304 is presented in Sects. 3 and 4 for the quiescent and variable emission, respectively. In Sect. 5 we focus on the neutrino output expected for the same blazar, analyzing the detectability with a next generation neutrino telescope such as KM3NeT. We finish in Sect. 6 with a discussion.


\section{Description of the model}

\begin{figure}[htp]
   \centering
   \includegraphics[trim = 10mm 80mm 0mm 40mm, clip,width=0.95\linewidth,angle=0]{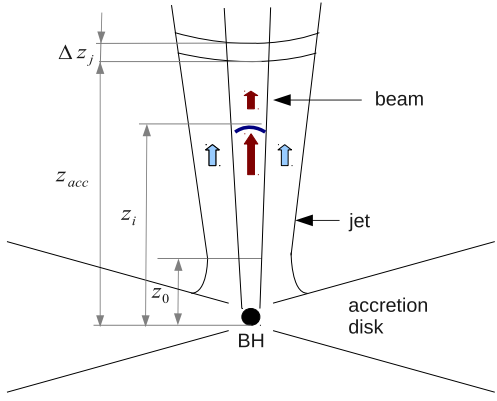}
   \caption{Basic elements of the model. See the text for details.}
              \label{FigSketch}%
    \end{figure}

We assume that matter is captured by the central black hole through a dissipationless accretion disk (Kelner \& Bogolobov 2010) and that a fraction of this accreted material is expelled by the accretion disk in two oppositely directed jets.
An inner beam of relativistic electrons and positrons is launched by the spinning black hole. This two-{component} setup is similar to the ones implemented in several previous models (e.g. Sol et al. 1989; Romero 1995; Ghisellini et al. 2005; Boutelier et al. 2008). A sketch of the basic elements  of the scenario is depicted in Fig. \ref{FigSketch}. 

{The introduction of a model with two components is motivated by observations as a means to reconciling available data with the unified AGN paradigm (e.g. Urry \& Padovani 1995), as noted by Chiaberge et al. (2000). One-component jet models require high bulk Lorentz factors ($10$--$20$) when applied to blazars and lower values when applied to FR radiogalaxies. Including of a second component serves to solve this discrepancy, for example, through the combination of a fast spine and a slower (but still relativistic) layer, so that the emission at small viewing angles is dominated by the fast component, whereas the slow component dominates for larger angles.} 

{Some additional observational facts also favor the two-component interpretation. For instance, the presence of a fast spine surrounded by a slower outflow can be inferred from the observed limb-brightened radio morphology of the radiogalaxies jets (Giroletti et al. 2004). On the other hand, whereas VLBI observations of powerful TeV BL Lacs suggest that the pc-scale jets move slowly (Piner \& Edwards 2004; Giroletti et al. 2004), rapid variability of the TeV emission implies that, in the region where this emission is produced, the jet should be extremely relativistic (e.g., Dondi \& Ghisellini 1995; Ghisellini et al. 2002; Konopelko et al. 2003).}

{On the theoretical side, the presence of both a spinning black hole and an accretion disk would unavoidably yield the launching of outflows through the BZ process (Blandford \& Znajek 1977) and the BP process (Blandford \& Payne 1982). The coexistence of both mechanisms has been investigated by Meier (2003), among others. Whereas the BZ process yields the launching of a Poynting flux that gives rise to a leptonic beam, the BP process can generate a baryon-rich jet launched centrifugally from the inner accretion disk (e.g. Komissarov et al. 2007; Sadowsky \& Sikora 2010).}

{The stability and structure of this type of jets have been discussed, for instance, in Hardee
(2007), Narayan et al. (2009), and Perucho (2012). For general discussion of fluid instabilies
see, e.g., Shore (2008).}


For the jet, we adopt the model developed by Reynoso et al. (2011). Equipartition between jet kinetic energy and magnetic energy takes place at $z_0= 50 R_{\rm g}$ from the black hole, the jet half-opening angle is $\xi_{\rm j}$, the jet Lorentz factor is $\Gamma_{\rm j}$ at $z_0$, and the viewing angle is $i_{\rm j}$. The bulk kinetic power of the jet at $z_0$ is a fraction $q_{\rm j}$ of the Eddington power,  
\beq 
\left.L_{\rm j}^{\rm(kin)}\right|_{z_0}= q_{\rm j}L_{\rm Edd}=(\Gamma_{\rm j}-1)\dot{m}_{\rm j} c ^2,
\eeq
and the cold particle density in the jet is 
\beq 
  n_{\rm j}= \frac{\dot{m}_{\rm j}}{m_p\pi z^2 \chi_{\rm j}^2 v_{\rm j} },\label{ncoldjet}
\eeq 
  with $\chi_{\rm j}=\tan{\xi_{\rm j}}$. The magnetic field varies along the jet according to $$ B(z)= B_0\left(\frac{z_0}{z}\right)^m \, \, {\rm with}\, m\in(1,2),$$ and the $\Gamma_{\rm j}$ increases slowly as magnetic energy density
decreases along the jet. 

The $e^+e^-$-beam, with a half-opening angle $\xi_{\rm b}< \xi_{\rm j}$, is confined by the jet by requiring that the pressure is less than in the jet:  $P_{\rm b}=\eta_{\rm b}P_{\rm j}$, where $\eta_{\rm b}<1$. The initial Lorentz factor of the beam is $\Gamma_{\rm b}^{(0)}$ at distance $z_{\rm b}^{(0)}< z_0$ from the black hole, where equipartition with the beam magnetic energy holds. This condition is used to fix $B_{\rm b}^{(0)}$. For $z>z_{\rm b}^{(0)}$ in the beam, we assume 
\beq 
 B_{\rm b}=B_{\rm b}^{(0)} \left(\frac{z_{\rm b}^{(0)}}{z}\right)^{m_{\rm b}} \,{\rm with} \, \, m_{\rm b}\in(1,2),
\eeq 
so that the Lorentz factor of the beam also increases gradually along it. This behavior of the magnetic field also enables Kelvin-Hemlholtz instabilities to develop, as we discuss below in Sect. \ref{SecVariable}, where we deal with the production of variable emission in the beam. As mentioned, this contribution is added to the one from the quiescent jet, which is discussed in next.

\section{The quiescence emission from the jet}
The quiescent state of the emission can be reproduced using a lepto-hadronic
model for the jet. Acceleration of relativistic electrons and protons takes place at $z_{\rm acc}>z_0$, where the power injected in these primary particles is $L_e$ and $L_p= a L_e$, with $a>0$. 
The total power in the injected, $L_{\rm rel}= L_e(1+a)$, is a fraction of the bulk kinetic power of the jet, $L_{\rm rel}= q_{\rm rel}L_{\rm j}^{(\rm kin)}$ (see Reynoso et al. 2011).
 The distribution of particles in the steady jet, $N(E,z)$ in units ${\rm GeV}^{-1}{\rm cm}^{-3}$, is governed by an inhomogeneous transport equation with cooling and convection: 
 \beq  \frac{\partial\left( \Gamma_{\rm j} v_{\rm j} N(E,z)\right)}{\partial z}+
\frac{\partial \left( b(E,z) N(E,z) \right) }{\partial E}+ \frac{N(E,z)}{T_{\rm
dec}(E) } = Q(E,z),\label{transporteq}
\eeq where $z$ is the distance to the black hole in the AGN frame, and $b(E,z)=-\frac{dE}{dt}$.
  \begin{figure*}[htp]
   \centering
   \includegraphics[trim = 10mm 0mm 0mm 0mm, clip,width=0.9\linewidth,angle=0]{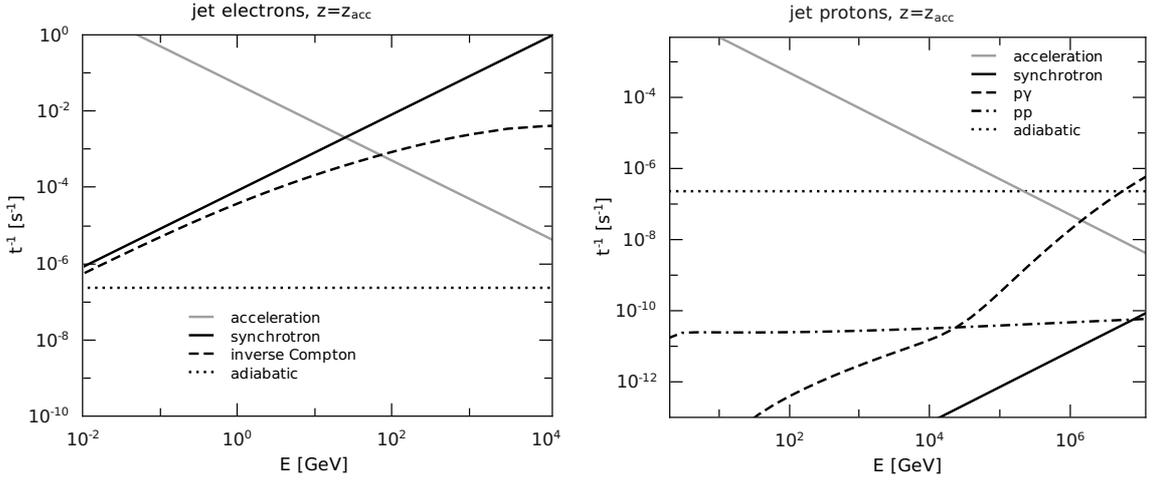}
   \caption{Acceleration and cooling rates for primary electrons and protons in the jet at $z_{\rm acc}$.}
              \label{Figcoolingjet}%
    \end{figure*}

This equation is solved in the jet co-moving frame, and the injection or source term is taken as
\beq	
  Q(E,z)= K_i\left(\frac{z_{\rm acc}}{z}\right)^2{E}^{-s}\exp{\left[-\left(\frac{E}{{E}^{({\rm max})}_{i}}\right)\right]},
\eeq
where the cut-off energy ${E}^{({\rm max})}_{i}$ is determined by equating the acceleration rate to the cooling one. 
The particle injection is normalized in the AGN frame, that where the central black hole is at rest, as
 \beq
 L_i=  \int_{\Delta E} dE'\int_{4\pi} d\Omega' \int_{\Delta V} dV \; E' \; \frac{d Q'_i(E',z)}{d\Omega'},
 \eeq
where 
\beq
\frac{d Q'_i(E',z)}{d\Omega'}= \left[\frac{E'^2- m_i^2c^4}{{E}^2- m_i^2c^4}\right]^{1/2} \frac{Q_i(E,z)}{4\pi},\label{Qjet2obs}
 \eeq
and 
 \beq
   E= \Gamma_{\rm j}\left(E'- \beta_{\rm j} \cos{\theta} \sqrt{{E'}^2-m_i^2c^4}\right).
 \eeq

The acceleration rate is given by
\beq
  {t}_{\rm acc}^{-1}(E,z) =  \eta \frac{c e B(z)}{E} \label{tacc},
 \eeq
with an efficiency $\eta$, and the loss rate corresponds to synchrotron emission, inverse Compton interactions, $pp$ and $p\gamma$ interactions, and adiabatic cooling (see Reynoso et al. 2011). In Fig. \ref{Figcoolingjet}, we show the acceleration and cooling rates for electrons and protons at the position $z_{\rm acc}$ in the jet. The list of assumed and derived parameters for the jet is shown in Table 1. {The values for the black
hole mass and redshift are taken as $M_{\rm bh}=10^9M_\odot$, and $z_{\rm rs} = 0.116$, respectively, following
Aharonian et al. (2007), and the column density of HI is taken as $N_{H}=1.3\times 10^{20}{\rm cm}^{-2}$ after
Lockman \& Savage (1995). The rest of the parameters in Table 1 corresponds to the above
described jet model, which is discussed in detail in Reynoso et al. (2011). These parameters
are fixed to obtain a SED that fits the quiescence emission represented by the same set of
observational data used in Aharonian et al. (2009).
}

%

   \begin{table}[!htp]
      \caption[]{Model parameters for the quiecence emission from the jet of PKS 2155-304}
         \label{Tab:paramsCena}
    \begin{minipage}{\textwidth}
    \begin{tabular}{|l|l|l|}
    \hline
        Parameter      &  \text{Description} & {\rm Value}   \\ \hline
            $M_{\rm bh}$& \text{black hole mass} & $10^{9} M_{\odot}$\\
            $q_{\rm j}$ & \text{ratio $2 L_{\rm j}^{\rm (kin)}/L_{\rm Edd}$}  & $0.1$ \\
            $\Gamma_{\rm j}(z_0)$ & \text{bulk Lorentz factor of the jet at $z_0$}  & $10$   \\
            $i_{\rm j}$& \text{viewing angle} &  $1^\circ$     \\
            $\xi_{\rm j}$ &\text{jet's half-opening angle} &  $1.5^\circ$ \\
            $q_{\rm rel}$ & \text{jet's content of relativistic particles} & $0.1$ \\
            $a$ & \text{hadron-to-lepton power ratio}     & $50$        \\
            $z_0$ & \text{jet's launching point}   & $50 \ R_{\rm g}$ \footnote{$R_{\rm g}=1.48\times 10^{14}\rm{cm}$}\\
            $q_{\rm m}$ & \text{magnetic to kinetic energy ratio at $z_{\rm acc}$} & $0.02$\\
            $z_{\rm acc}$ & \text{injection point} & $575 \ R_{\rm g}$  \\
            $\Delta z_{\rm j}$& \text{size of injection zone} & $z_{\rm acc} \tan\xi_{\rm j}$  \\ 
            $m$ & \text{index for magnetic field } & $1.8$\\
            $s$ & \text{spectral index injection} & $2.1$ \\
            $\eta$& \text{acceleration efficiency} & $10^{-6}$\\
            $E_p^{\rm (min)}$& \text{minimum proton energy} & $2 \,{\rm GeV}$ \\
            $E_e^{\rm (min)}$& \text{ minimum electron energy} & $0.9 \,{\rm GeV}$ \\
            $N_H$& \text{column density of HI} & $ 1.3\times 10^{20} \,{\rm cm^{-2}}$\\
          $z_{\rm rs}$& \text{redshift} & $0.116$ \\
         \hline
    \end{tabular}
  \end{minipage}  
   \end{table}

%
%

\begin{figure}
   \centering
\includegraphics[trim = 0mm 0mm 0mm 0mm, clip,width=0.9\linewidth]{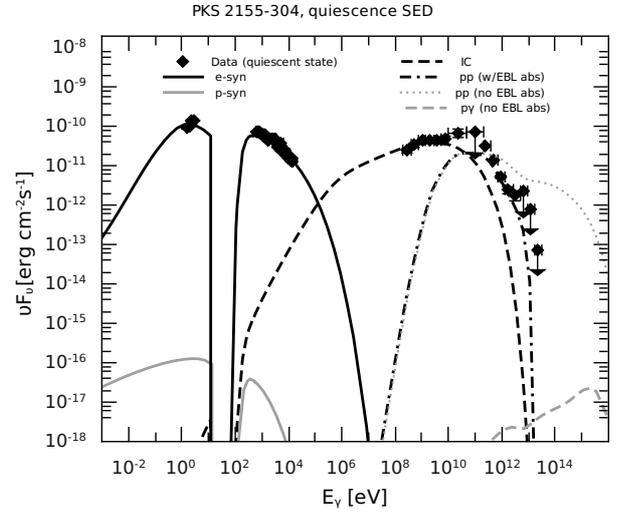} 
      \caption{SED corresponding to the quiescent state of PKS 2155-304. The contributions shown are electron synchrotron (black solid line), IC (dashed line), $pp$ (dashed-dotted line), proton synchrotron (gray solid line), and $p\gamma$ (gray dashed line).  
              }
         \label{Figsedjet}
   \end{figure}

The SED is dominated by synchrotron and inverse Compton (IC) emission of the electrons, and $pp$ orginated VHE gamma-rays, as shown in Fig. \ref{Figsedjet}. In our calculations, the VHE energy emission has been corrected by the effect of EBL absorption following Dominguez et al. (2010).
%

\section{Variable emission from the beam}\label{SecVariable}

\begin{figure}
   \centering
\includegraphics[trim = 0mm 0mm 0mm 0mm, clip,width=0.95\linewidth]{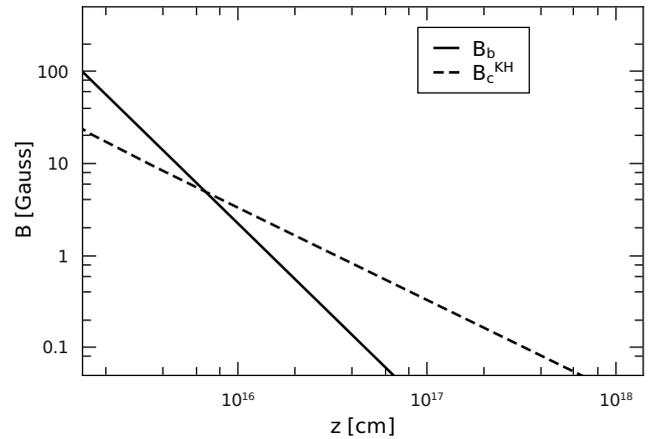} 
      \caption{Magnetic field in the beam (red line) and critical magnetic field below which KH instabilities can develop.}
         \label{FigBbm}
   \end{figure}

The variable emission is assumed here to have its origin in shocked regions in the internal $e^+e^-$ beam that can arise, e.g., due to Kelving-Helmholtz (KH) instabilities. These instabilities will be able to develop in the beam if the magnetic field is below a critical value (e.g. Romero 1995):
\beq
B_{\rm b}< B_{c}^{\rm KH}= \frac{\sqrt{4\pi n_{\rm b}m_ec^2(\Gamma_{\rm b}^2-1)}}{\Gamma_{\rm b}}.
\eeq
In Fig. \ref{FigBbm} we show the beam magnetic field and the critical value $B_{c}^{\rm KH}$, considering that $\Gamma_{\rm b}^{\rm (0)}=5$ at $z_{\rm b}^{\rm(0)}= 5 R_{\rm g}$. It can be seen from this plot that for $z\gtrsim 10^{16}{\rm cm}$ KH instabilities start to develop and create inhomogeneities if the pinching modes dominate (see Romero 1995). The rarefactions produced in the beam by the instabilities act as obstacles for the fast plasma, and strong shocks are expected to appear at $z\sim 10^{16}$cm. These shocks reaccelerate the pairs (Araudo et al. 2010) producing a local injection of relativistic particles. 

A shock front originating in the beam at a position $z_{i}$ with a Lorentz factor $\Gamma
_{\rm s}$ will form a post-shocked region of a certain size 
\beq\Delta z= \dfrac{z_i {\chi_{\rm b} } } {b},
\eeq
 with $\chi_{\rm b}=\tan{\xi_{\rm b}}$, and particles are injected during a time 
\beq \Delta t_{\rm inj, \, obs} \approx  \frac{\Delta z}{c},
\eeq as seen in the observer frame.

The distribution of the injected pairs is computed in the post-shock frame, where the population is assumed to be isotropic. The corresponding source term is a power law with an exponential cutoff at the maximum Lorentz factor $\gamma_{\rm max}$:
\beq
 q_e(\gamma,t)= H(t-t_{\rm on})H(t_{\rm off}-t) \ q_0 \ \gamma^{-s_{\rm b}}\exp{\left(-\frac{\gamma}{\gamma_{\rm max}}\right)},
\eeq  is 
where $H$ is the Heavyside step function, {$t_{\rm on}$ the time when the injection is switched on, and the time when it is switched off is}
\beq
t_{\rm off}= t_{\rm on}+ \frac{\Delta t_{\rm inj, \, obs}}{\Gamma_{\rm ps}(1+ z_{\rm rs})},
\eeq
{both corresponding to the post-shock frame. We note that, for example, the onset time is transformed to the observer frame as}
\beq
  t_{\rm on,\, obs}= \Gamma_{\rm ps}(1- z_{\rm rs}) t_{\rm on}.
\eeq

The constant $q_0$ is found through normalization taking into account the particle density in the post-shock frame:
 \beq
    n_{\rm ps}=  \int_{t_{\rm on}}^{t_{\rm off}} dt \int_{\gamma_{\rm min}}^{\infty} d\gamma \  q_e(\gamma,t).
 \eeq
This can be related to the beam density $n_{\rm b}$ by the compresion factor $\zeta$ as 
\beq
 n_{\rm ps}= \zeta n_{\rm b}, \ \mbox{with} \ \ \zeta=\left(\frac{\hat{\gamma}\Gamma'_{\rm ps}+1}{\hat{\gamma}-1}\right), 
\eeq
where $\hat{\gamma}=4/3$ is the polytropic index, $$\Gamma'_{\rm ps}= \Gamma_{\rm ps}\Gamma_{\rm b}(1-\beta_{\rm ps}\beta_{\rm b})$$ is the Lorentz factor of the shock in the undisturbed beam frame, and the Lorentz factor of the post-shock (ps) region is $\gamma_{\rm ps}\approx \gamma_{\rm s}/ \sqrt{2}.$ The density of the beam, in turn, is fixed as a fraction $\lambda$ of the jet density of Eq. \ref{ncoldjet}:
 \beq
   n_{\rm b}= \lambda n_{\rm j}.	
 \eeq

The maximum Lorentz factor of the locally accelerated pairs $\gamma_{\rm max}$ is obtained from the balance between the acceleration rate and the cooling rate, where the latter is due to synchrotron and synchrotron self-Compton (SSC) interactions.

We solve the following kinetic equation to obtain the distribution of pairs as a function of time $n_e(\gamma, t):$
\beq
 \frac{\partial  n_e}{\partial t}- \frac{ \partial}{\partial \gamma}  \left( n_e \ \left|\dot{\gamma}\right|_{\rm loss}  \right) = q_e(\gamma,t),  
\eeq
where the loss term is given by 
\beq
 \left|\dot{\gamma}\right|_{\rm loss}=\left|\dot{\gamma}\right|_{\rm syn}+ \left|\dot{\gamma}\right|_{\rm SST}.
\eeq
An iterative process is implemented to work out $n_e(\gamma,t)$ from this equation. As a zero-order approximation, we adopt the solution corresponding to no SSC interactions, just synchrotron losses. Then, the solution is succesively improved by computing the SSC cooling term using the previus solutions found for $n_e(\gamma,t)$. This cooling rate is computed in the Thompson regime to save computing time, since this simplification was tested not to affect the final result. The self-synchrotron Thomson (SST) cooling rate is then (e.g. Schlickeiser \& Lerche 2007):
 \beq
 |\dot{\gamma}|_{\rm SST}=  A_0 \gamma^2 \int_1^{\infty}d\gamma \gamma^2 n_e(\gamma,t),  
 \eeq
where 
$$A_0= \frac{4}{3}c\sigma_{\rm T}\frac{B_{\rm ps}^2}{8\pi}$$

The synchrotron cooling rate is 
 \beq
   |\dot{\gamma}|_{\rm syn}= \frac{ \sigma_{\rm T} B_{\rm ps}^2 }{6\pi m_ec}\gamma^2.
 \eeq
The acceleration rate is worked out using Eq. (\ref{tacc}) along with the magnetic field corresponding to the
post-shock frame,

\beq
  B_{\rm ps}(z_i)= B_{\rm b}(z_i) \sqrt{\zeta^2+1}.
\eeq

For any injection point $z_i$ in the beam, we solve the kinetic equation, and work out the resulting synchrotron and SSC emission. 
{In the local AGN frame, the emissivity is given by}
 \beq
\frac{dQ'_\gamma(E',t')}{d\Omega'}= D \ {\frac{dQ_\gamma(E,t)}{d\Omega}},\label{Qgtrans}
 \eeq
{where the Doppler factor is $D=\Gamma_{\rm ps}^{-1}(1-\beta_{\rm ps}\cos i_{\rm j})^{-1}$, $E$ is the photon energy in the ps frame, and $ E'= D\ E$ is the photon energy in the local AGN frame.}
In the case of synchrotron radiation, the emissivity in the ps-frame is
\be
\frac{dQ_{\gamma,{\rm syn}}(E,t)}{d\Omega}= \frac{1}{4\pi E}{\left(\frac{1- e^{-\tau_{\rm SSA}(E,t)}}{\tau_{\rm SSA}(E,t)} \right)}\int_{m_e c^2}^{\infty} d\gamma' P_{\rm syn}
    n_{e}(\gamma',t),\label{Qsyn}
\ee
where $\tau_{\rm SSA}$ is the optical depth corresponding to synchrotron-self absorption, and the synchrotron power per unit energy emitted by the electrons is given by (Blumenthal \& Gould 1970):
  \be
 P_{\rm syn}(E,\gamma,z)= \frac{\sqrt{2} e^3 B_{\rm ps}}{m_e c^2 h}
\frac{E}{E_{\rm cr}} \int_{E/E_{\rm cr}}^\infty d\zeta
K_{5/3}(\zeta), \ee
 where $K_{5/3}$ is the modified Bessel function of order $5/3$ and
$$E_{\rm cr}= \frac{\sqrt{6}he B_{\rm ps}}{4\pi m_e c}\gamma^2.$$ 

The soft photon density in the ps frame is
\beq
 n_{\rm ph}(E,t)\approx {\ Q_{\rm syn}(E,t)}\frac{z_i\tan\xi_{\rm b}}{c},
\eeq
and the IC emissivity in the this frame is
 \begin{multline}
Q_{\gamma, {\rm IC}}(E,t)= \frac{r_e^2 c}{2}\int_{{E}_{\rm(min)}}^{E} dE_{\rm ph} \frac{n_{\rm ph}(E_{\rm ph},t)}{E_{\rm ph}} \\
  \times \int_{\gamma_{\rm min}}^{\gamma_{\rm max}}d{\gamma'} \frac{n_e({\gamma'},t)}{\gamma^2}F(q).
  \end{multline}
Here {$r_e$ is the classical electron radius, and} we integrate in the target photon energy $E_{\rm ph}$ 
and in the electron Lorentz factors $\gamma'$ between
 \beq
 \gamma_{\rm min}= \frac{E}{2 m_e c^2}+\sqrt{\frac{E}{4E_{\rm ph}}+\frac{{E}^2}{4m_e^2c^4}}
\  \mbox{ and}  \  \gamma_{\rm max}= \frac{E}{m_ec^2(1-\frac{E}{E_{\rm ph}})}.\nonumber
  \eeq
The function $F(q)$ is given by
\be
  F(q)= 2q \ln q + (1 + 2q) (1 - q)+ \frac{1}{2}(1 - q)\frac{(q \Gamma'_{\rm e})^2}{1 + \Gamma'_{\rm e} },
\ee
with $\Gamma_{\rm e}=4 E_{\rm ph} \gamma'/(m_ec^2)$ and
 $$q=\frac{E}{\Gamma_{\rm e}E_{\rm ph}(1-{E/}{E_{\rm ph}})}. $$
  
The differential photon flux is first obtained in the AGN local frame as
\beq
 \frac{d\Phi'_{\gamma}(E',t')}{dE'}= (\pi z_i^2\tan^2\xi_{\rm b} \Delta z) \frac{d Q'_\gamma(E',t') } {d\Omega'} \frac{d\Omega'} {dA'}  \exp{\left[-\tau_{\gamma\gamma}\right]},
\eeq  
where $\tau_{\gamma\gamma}= \tau_{\gamma\gamma}^{\rm (b)}+ \tau_{\gamma\gamma}^{\rm (j)}$ is the $\gamma\gamma$ optical depth due to the jet and the beam photons. The latter is estimated as
 \begin{multline}
 \tau_{\gamma\gamma}^{\rm (b)}(E',t')= \left(z_i\tan\xi_{\rm b}\right) \int_{-1}^1dx(1-x) \\ \times \int_{\frac{2m_e^2c^4}{E(1-x')}}^{\infty} dE_{\rm ph} n_{\rm ph}(E_{\rm ph},t') \sigma_{\gamma\gamma}(E,E_{\rm ph}),
 \end{multline}
and $\tau_{\gamma\gamma}^{\rm (j)}$ is obtained from a similar expression.

\begin{figure*}
   \centering
\includegraphics[trim = 0mm 0mm 2mm 0mm, clip,width=0.92\linewidth]{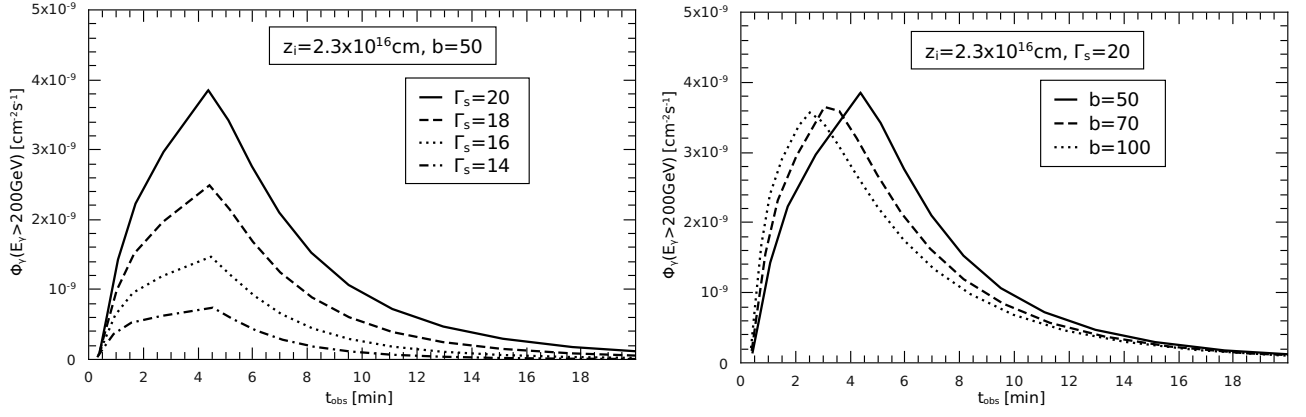} 
      \caption{Light curves of gamma-rays above $200$ GeV produced by $e^+e^-$ injection in the beam at $z_i= 2.3\times 10^{16}$cm with $\Gamma_{\rm s}=\{14,16,18,20\}$ in the left panel, and with $\Gamma_{\rm s}=20$ and $b=\{50,70,100\}$ in the right panel.  
              }
         \label{FigLc2panels}
   \end{figure*}

\begin{figure*}
   \centering
\includegraphics[trim = 0mm 0mm 2mm 0mm, clip,width=0.92\linewidth]{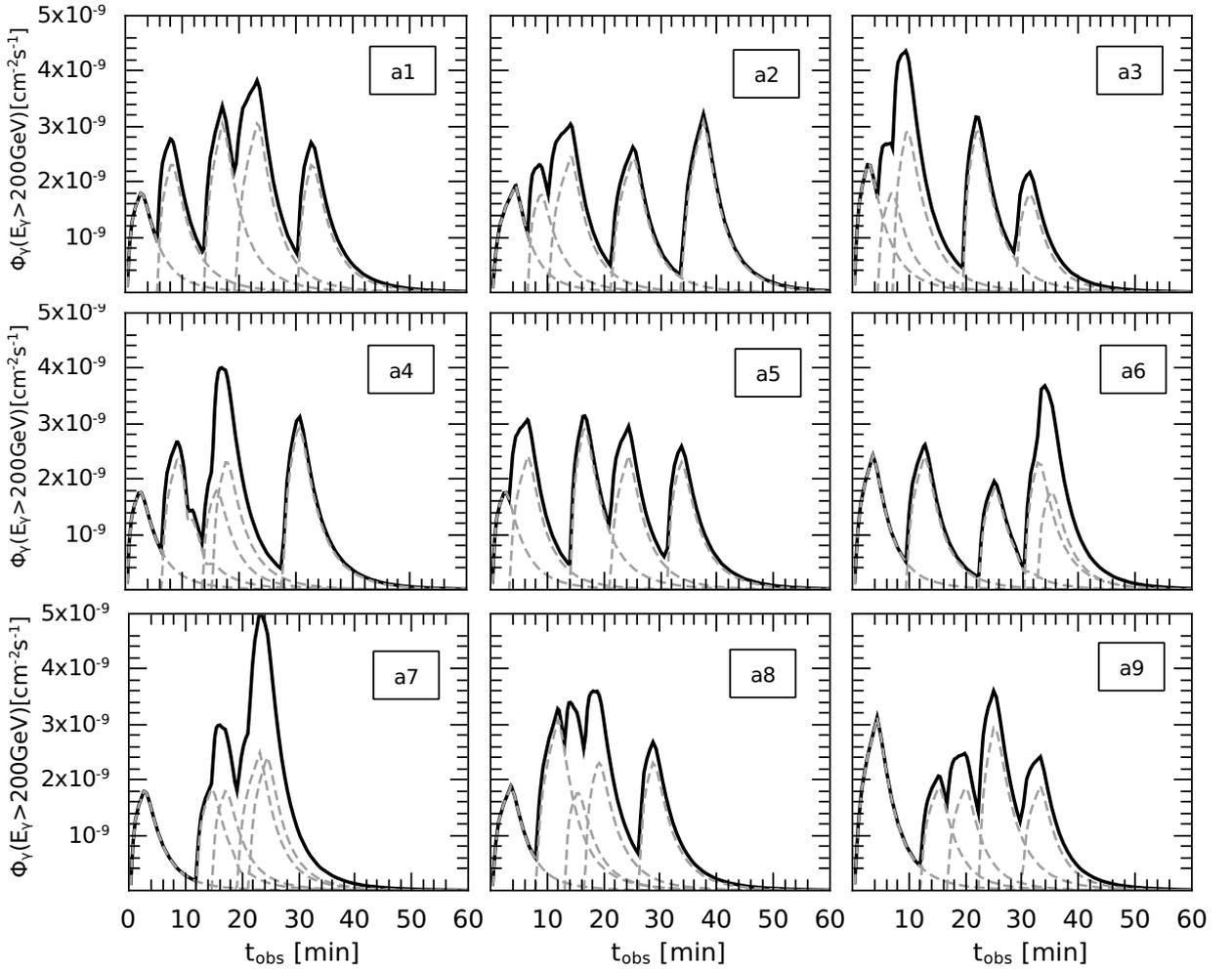} 
      \caption {Light curves of gamma-rays above $200$ GeV produced by $e^+e^-$ injection in the beam at $z_i= 2.3\times 10^{16}$cm with the randomly chosen values for $\Gamma_{\rm s}$, $b$, and $t_{\rm on}$ of Table \ref{tabrandomzin0}.}\label{Figrandomzin0}
   \end{figure*}

If $d_{\rm C}$ is the comoving distance, then
$$ \frac{d\Omega'}{dA'}= \frac{1}{d_{\rm C}^2},$$
and taking into account that the observed energy and time are transformed as
 $$ E_\gamma= \frac{E'}{1+z_{\rm rs}} \ \mbox{and} \ t_{\rm obs}=t'(1+z_{\rm rs}),$$
the differential photon flux arriving on Earth from a source at redshift $z_{\rm rs}$ is
\begin{multline}
\frac{d\Phi_{\gamma} ( E_\gamma,t_{\rm obs} ) } {dE_{\gamma}} = \frac{(1+z_{\rm rs})^2}{d_{\rm L}^2} \frac{ d \Phi'_\gamma \left( E_\gamma ( 1+z_{\rm rs} ),\frac{ t_{\rm obs} } { 1+z_{\rm rs} } \right)} { dE' }  \\
\times \exp{ -\left[\tau_{\rm EBL} (E_\gamma,z_{\rm rs})\right] },
  \label{dgammaflux}
\end{multline}  
where $d_{\rm L}$ is the luminosity distance and $\tau_{\rm EBL}$ is the optical depth due to $\gamma\gamma$ absorption on the extragalactic background light. In these terms, the usual amount $\nu F_\nu$ in units ${\rm erg \ cm}^{ -2 }{\rm s}^{ -1 }$ is expressed as $E_\gamma^2\frac{ d\Phi_\gamma } { dE_\gamma}$. A light curve can be obtained simply by integrating Eq. (\ref{dgammaflux}) on the desired energy range for different times. In Fig. \ref{FigLc2panels} we show the light curves that can be obtained if the injection point in the beam is $z_i=2.3\times 10^{16}{\rm cm}$ for different shock Lorentz factors $\Gamma_{\rm s}$ and sizes of the injection zone $\Delta z= R_{\rm b}/b$. The values adopted for beam parameters are shown in Table \ref{tablebeam}. More complex light curves can be obtained, for instance, by succesive shock injections at the same position with $\Gamma_{\rm s}$, $b$, and $t_{\rm on}$ chosen randomly in the ranges:
 \be
 \Gamma_{\rm s} & \in & \{ 17,18,19 \} \\
 b &\in& \{50,60,70,80,90\} \\
 t_{\rm on} &\in& [1..15]\rm{min}.
  \ee
Some example light curves obtained in this way are plotted in Fig. \ref{Figrandomzin0} with the parameters shown in Table \ref{tabrandomzin0}. 

\begin{table}
  \caption[]{Beam parameters}\label{tablebeam}
  \begin{minipage}{\textwidth}
    \begin{tabular}{|l|l|l|}
    \hline
      Parameter      &  \text{Description} & {\rm Value}   \\ \hline
      $z_{\rm(b)}^{\rm (0)}$ &  launching point [$R_{\rm g}$]& $5$\\
     $\Gamma^{\rm (0)}_{\rm b}$ & Lorentz factor at $z_{\rm(b)}^{\rm (0)}$ & $5$\\
     $\lambda$ & beam to jet density ratio & $0.1$\\
     $\xi_{\rm b}$ & half-opening angle $[^\circ]$ & $ 0.5$ \\        
       \hline
    \end{tabular}
  \end{minipage}
\end{table}

\begin{table}
  \caption[]{Randomly chosen values for the parameters corresponding to Fig.\ref{Figrandomzin0} }\label{tabrandomzin0}
  \begin{minipage}{\textwidth}
    \begin{tabular}{|l|l|l|l|}
    \hline
      Label & $\Gamma_{\rm s}$ & $b$ & $t_{\rm on} {\rm [min]}$  \\ \hline
      $a1$ & $17,18,19,19,18$ & $80,80,60,50,80 $ & $ 5.3,8.1,5.5,11 $ \\
       $a2$ & $17,17,18,18,19$ & $50,90,50,50,50 $ & $ 6.3,3.6,11,12.4 $ \\
       $a3$ & $18,17,19,19,17$ & $90,80,80,80,80 $ & $ 4.2,2.6,12.3,9.4 $ \\
       $a4$ & $17,18,17,18,19$ & $90,70,80,90,70 $ & $ 6, 7.2,1.7,12.3 $ \\        
       $a5$ & $17,18,19,18,18$ & $90,60,80,60,90 $ & $ 3, 10.7, 7, 10.1 $ \\                
       $a6$ & $18,18,17,18,17$ & $60,60,70,90,80 $ & $ 9.1, 12.9, 8.1, 2.2 $\\
       $a7$ & $17,17,17,18,18$ & $80,70,80,50,60 $ & $ 11.6, 2.8, 4.4, 2.1 $\\                               
       $a8$ & $17,19,17,18,18$ & $80,70,80,50,60 $ & $ 7.6, 5.2, 3.5, 9.6 $\\                               
       $a9$ & $19,17,17,19,17$ & $50,60,60,70,60 $ & $ 11.7, 4.7, 5.5, 7.6 $\\                                      
               \hline
    \end{tabular}
  \end{minipage}
   
\end{table}

\begin{figure}
   \centering
\includegraphics[trim = 0mm 0mm 0mm 0mm, clip,width=0.9\linewidth]{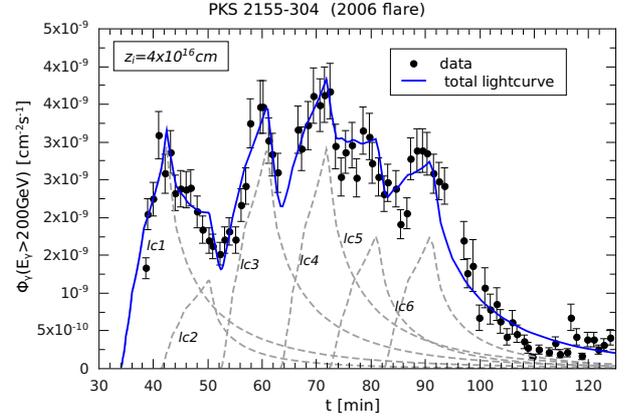} 
      \caption{Lightcurve of gamma-rays above $200$ GeV (blue line), generated as the superposition of six injections in the beam ({\it lc1,lc2,lc3,lc4,lc5,lc6}, in gray-dashed lines), as compared to data of the 2006 flare of PKS 2155-304. }\label{lctuned}
     \end{figure}

In particular, in view of the flaring activity detected in 2006 (Aharonian et al. 2007), we see that a simillar lightcurve can be obtained for the flux of gamma-rays above $200$ GeV by adding up six different injections at a fixed position in the beam, $z_i=4\times 10^{16}$cm, each being switched on at apropriate times $t_{\rm on}$, as seen in Fig. \ref{lctuned}. The values of the different parameters are shown in Table \ref{tablebeam}. The multiple shock scenario resembles the one proposed to explain the rapid variability in gamma-ray bursts (GRBs), e.g., Kobayashi et al. (1997). The whole SED evolution shown in Fig. 8 shows that multiwavelength observations, including studies of rapid variability in radio, X-ray, and $\gamma$-ray bands, can be used to test the proposed model.

\begin{figure}
   \centering
\includegraphics[trim = 0mm 0mm 0mm 0mm, clip,width=0.9\linewidth]{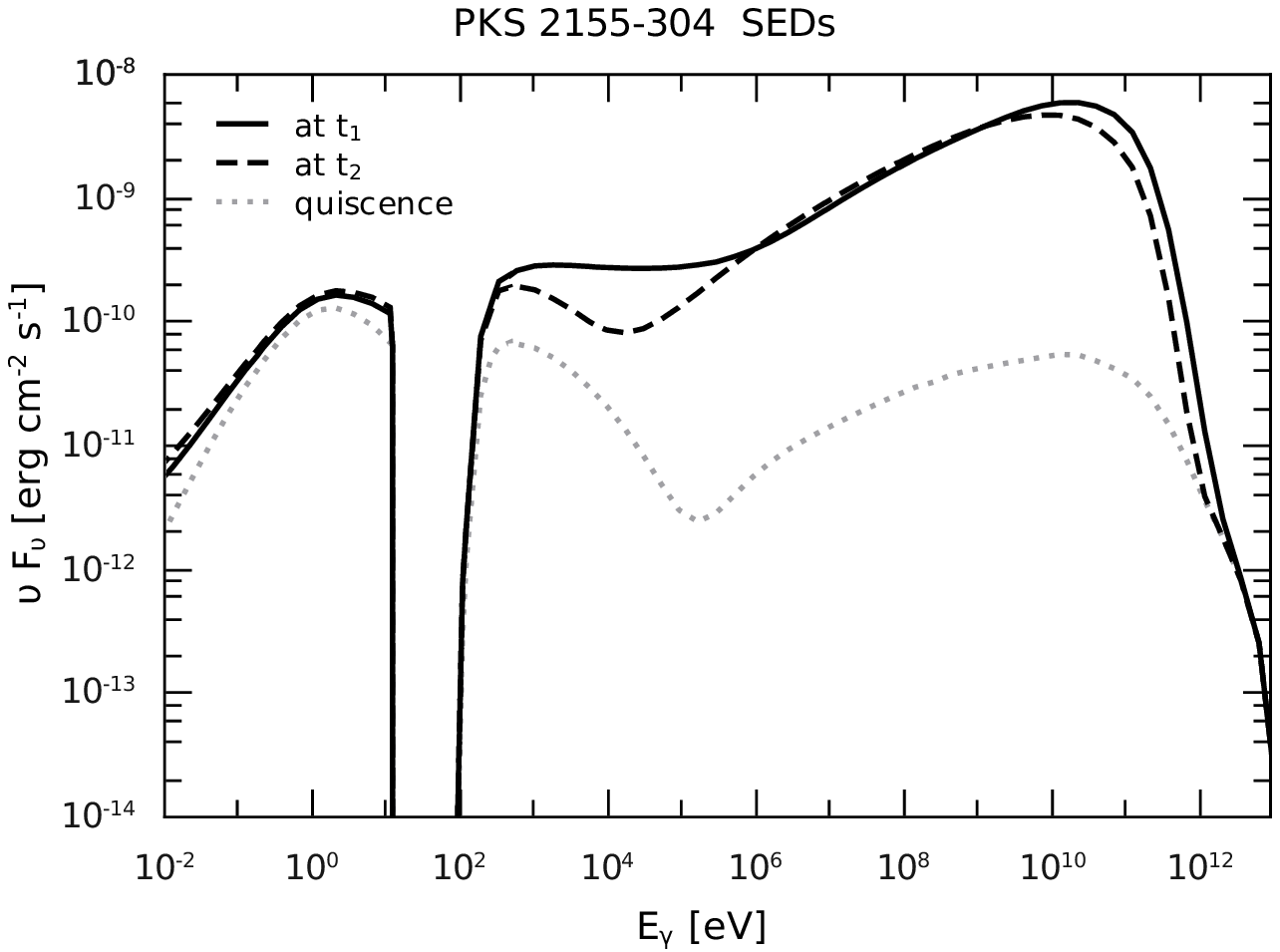}
      \caption{Instantaneus SEDs originating in the beam at $t_1\simeq 40 $min, and $t_2\simeq 50$min corresponding to the plot of Fig.\ref{lctuned}. The quiescent SED is also shown (gray dashed lines).}\label{instantseds}
     \end{figure}

\begin{figure}
   \centering
\includegraphics[trim = 0mm 0mm 0mm 0mm, clip,width=0.9\linewidth]{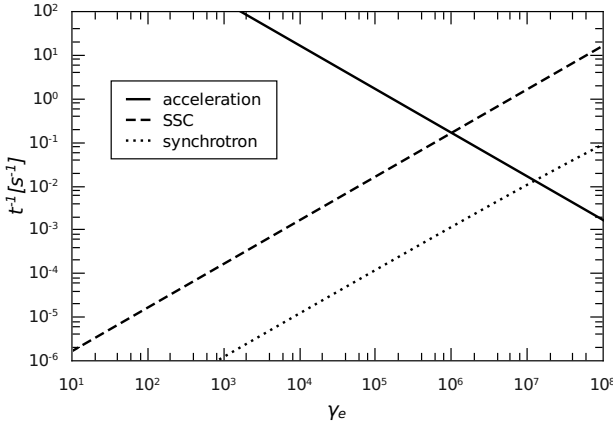}
      \caption{Acceleration and cooling rates {in the post-shock frame} corresponding to $t_1\simeq 40 $min in the lightcurve of Fig.\ref{lctuned}. }\label{figratest1}
     \end{figure}

In Fig. \ref{instantseds}, we show for illustration two instantaneus contributions to the SED due to the shocked beam at $t_1\simeq 40 $min and $t_2\simeq 50$min.
{The acceleration and cooling rates corresponding to $t_1$ are shown in Fig. \ref{figratest1}, where it can be seen that the Lorentz factors of the electrons are $\sim 10^6$ in the post-shock frame. The inverse Compton interactions with target photons created in the jet was verified to be unimportant for the adopted value of $z_i$.}

\begin{table*}
  \caption[]{Model parameters for the variable emission from the beam of PKS 2155-304}\label{tablebeam}
  \begin{minipage}{\textwidth}
    \begin{tabular}{|l|l|l|l|l|l|l|l|}
        \hline
        Parameter & description & lc1 value & lc2 value & lc3 value & lc4 value & lc5 value & lc6 value \\ \hline
        $\Gamma_{\rm s}$ & shock Lorentz factor & 13 & 10.5 & 13 & 13 & 11.5 & 11.5\\ 
        $\eta_{\rm b}$ & acceleration efficiency & $10^{-2}$ & $10^{-2}$ & $10^{-2}$ & $10^{-2}$ & $10^{-2}$ & $10^{-2}$ \\ 
      $s_{\rm b}$   & power law injection index & $2.1$ & $2.1$ & $2.1$ & $2.1$ & $2.1$ & $2.1$\\ 
      $z_i$ & position of injection at the beam [cm] & $4\times 10^{16}$ &  $4\times 10^{16}$ & $4\times 10^{16}$ & $4\times 10^{16}$ & $4\times 10^{16}$ & $4\times 10^{16}$\\
      $\Delta z$ & size of injection zone & $z_i {\chi_{\rm b}}/50$ & $z_i {\chi_{\rm b}}/50$ & $z_i {\chi_{\rm b}}/50$ & $z_i {\chi_{\rm b}}/50$ & $z_i {\chi_{\rm b}}/50$& $z_i {\chi_{\rm b}}/50$ \\
       $t_{\rm on,\,\rm obs}$ & observed onset time [min] & $0 $ & $7.5$ & $11$ & $11$ & $9$ & $10$ \\  
     $\gamma_{\rm min}$ & minimum Lorentz factor of injection & $ 10$ & $ 10$ & $ 10$ & $ 10$ & $ 10$ & $ 10$\\
     $B_{\rm ps}$ & post-shock magnetic field at $z_i$ [Gauss]& $ 0.95$ & $ 0.97$ & $ 0.95$ & $0.95$ & $ 0.96$ & $ 0.96$\\
        \hline
    \end{tabular}
  \end{minipage}

\end{table*}

\section{Neutrino emission}

The acceleration of protons in the jet and their interactions with synchrotron photons and cold jet protons lead to the production of VHE gamma-rays after $\pi^0$ decays, as seen in the SED of Fig. \ref{Figsedjet}. These $p\gamma$ and $pp$ interactions also give rise to charged pions that decay to muons and neutrinos. The stationary distributions of pions and muons in the jet, $N_\pi(E,z)$ and $N_\mu(E,z)$, are found using the transport equation Eq. \ref{transporteq}. The corresponding neutrino emissivity from direct pion decay can be computed in the jet comoving frame following Lipari et al. (2007):
\begin{multline}
Q_{\pi\rightarrow\nu_\mu}(E,z)= \int_{E}^{\infty}dE_\pi
T^{-1}_{\pi,\rm d}(E_\pi)N_\pi(E_\pi,z) \\ \times
\frac{\Theta(1-r_\pi-x)}{E_\pi(1-r_\pi)},
\end{multline}
with $x=E/E_\pi$ and $T_{\pi,\rm d}=2.6\times 10^{-8}{\rm s}$. 
The contribution from muon decays ($\mu^- \rightarrow e^- \bar{\nu}_e \nu_\mu$,  $\mu^+ \rightarrow e^+ {\nu}_e \bar{\nu}_\mu$) can be calculated as
\begin{multline}
Q_{\mu\rightarrow\nu_\mu}(E,z)= \sum_{i=1}^4\int_{E}^{\infty}\frac{dE_\mu}{E_\mu} T^{-1}_{\mu,\rm
d}(E_\mu)N_{\mu_i}(E_\mu,z) \\ \times \left[\frac{5}{3}-
3x^2+\frac{4}{3}x^3 
\right].
\end{multline}
In this expression, $x=E/E_\mu$, $\mu_{1,2}=\mu^{-,+}_L$, $T_{\mu,\rm d}=2.2\times 10^{-6}{\rm s}$, and
$\mu_{3,4}=\mu^{-,+}_R$. 

\begin{figure}
   \centering
\includegraphics[trim = 0mm 0mm 0mm 0mm, clip,width=0.7\linewidth]{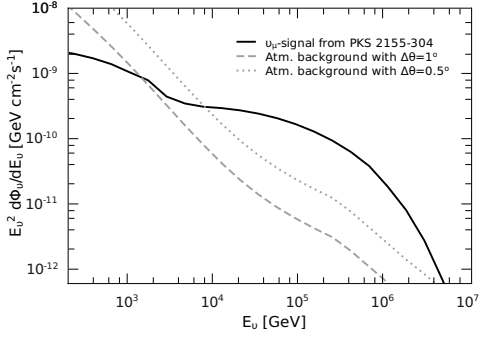}
      \caption{Predicted muon neutrino differential flux weighted by the squared neutrino energy as compared to the flux of atmospheric neutrinos for $\Delta\theta_{\rm bin}=\{0.5^\circ,1^\circ\}$. }\label{FigE2numu}
     \end{figure}

In the local AGN frame, the differential neutrino emissivity, in units (${\rm GeV}^{-1}{\rm cm}^{-3}{\rm sr}^{-1}{\rm s}^{-1}$)
is given by
\be
\frac{dQ'_\nu(E',z)}{d\Omega'}= D\frac{Q_{\pi \rightarrow \nu_\mu}(E,z) + Q_{\mu \rightarrow \nu_\mu}(E,z)}{4\pi},
\ee
where $z$ is the distance from the black hole to the position in the jet in the AGN frame and $D=\Gamma_{\rm j}^{-1}(1-\beta_{\rm j}\cos i_{\rm j})^{-1}$. The differential neutrino flux is then
\be
\frac{d\Phi'_\nu}{dE'}= \frac{1}{{d_{\rm C}^2}}\int_{z'>z_{\rm acc}} {dV'} \frac{dQ'_\nu(E',z')}{d\Omega'},
\ee
and in the observer frame, since the redshifted neutrino energy is $E_\nu= E'/(1+z_{\rm rs})$, 
\be
\frac{d\Phi_\nu(E_\nu)}{dE_\nu }= \frac{d\Phi'_\nu\left(E_\nu(1+z_{\rm rs})\right)}{dE'}.
\ee
On Earth, neutrino telescopes are used to observe this flux in the presence of the atmospheric neutrino background,
$\frac{d\Phi_\nu(E_\nu)}{dE_\nu d\Omega}$, which introduces more events depending on the size of the angular search bin of the detector, $$\Delta \Omega_{\rm bin}= 2\pi\left(1-\cos{\frac{\Delta\theta_{\rm bin}}{2}}\right).$$ This depends on the resolution of the detector $\Delta \theta_{\rm bin}$, which is expected to be below $1^\circ$ for KM3NeT.
In the case of PKS 2155-304, we compare in Fig.\ref{FigE2numu} the predicted neutrino flux with the atmospheric background for $\Delta\theta_{\rm bin}=\{0.5^\circ,1^\circ\}$.

\begin{figure}
   \centering
\includegraphics[trim = 0mm 0mm 0mm 0mm, clip,width=0.9\linewidth]{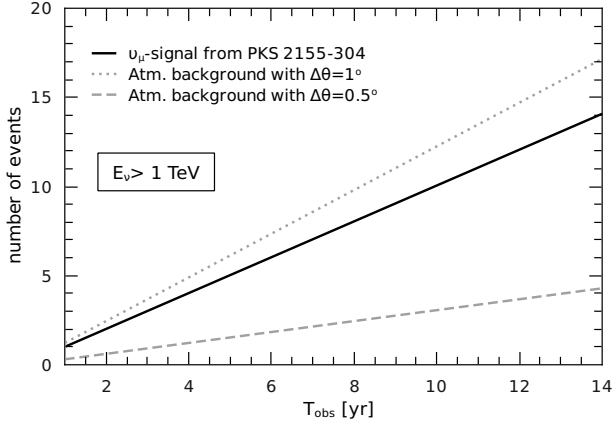}
      \caption{Muon neutrino events from PKS 2155-304 and atmospheric background for $\Delta \theta_{\rm bin}=\{0.5^\circ,1^\circ\}$ as a function of the observation time. }\label{eventsnu}
     \end{figure}

The signal to be searched for in neutrino telescopes such as ANTARES or KM3NeT, can be calculated as
 \beq
 N_{\rm s}(E_\nu> E_\nu^{\rm (min)})= T_{\rm obs} \int_{E_\nu^{\rm (min)}}dE_\nu \frac{d\Phi_\nu(E_\nu)}{dE_\nu } A_\nu^{\rm eff}(E_\nu),\label{Nsignal}
 \eeq  
where $A_\nu^{\rm eff}(E_\nu)$ is the neutrino effective area of the detector and $T_{\rm obs}$ is the observation time.
These signal events are to be detected among the background events caused by the flux of atmospheric neutrinos, which are given by
 \beq
 N_{\rm b}(E_\nu> E_\nu^{\rm (min)})= T_{\rm obs} \Delta\Omega_{\rm bin}\int_{E_\nu^{\rm (min)}}dE_\nu \frac{d\Phi^{\rm (atm)}_\nu(E_\nu)}{dE_\nu d\Omega } A_\nu^{\rm eff}(E_\nu).
 \eeq  
 Considering the neutrino effective area expected for this detector, we compute the predicted number of signal events of $(\nu_\mu+ \bar{\nu}_\mu)$ from PKS 2155-304 as a function of the observation time, and compare it with the background events corresponiding to $\Delta\theta_{\rm bin}=1^\circ$ and $0.5^\circ$. The result is shown in Fig. \ref{eventsnu} and corresponds to neutrinos above $E_\nu^{\rm (min)}=1$ TeV, yielding many events in a few years. 
 
\begin{figure}
   \centering
\includegraphics[trim = 0mm 0mm 0mm 0mm, clip,width=0.9\linewidth]{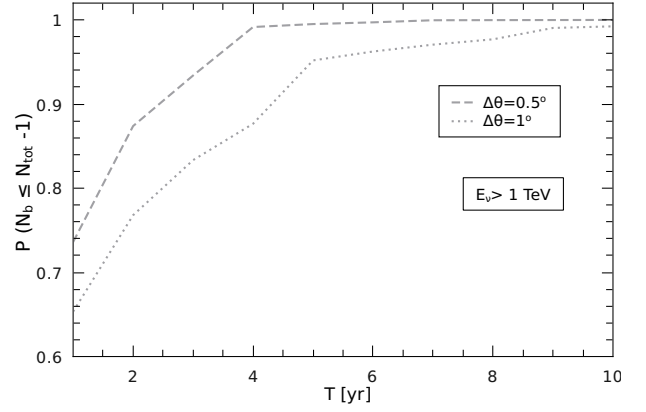}
      \caption{Cumulative probability that at least one event corresponds to the neutrino signal from PKS 2155-304 as a function of the observation time. }\label{FigCDF}
     \end{figure}

We can proceed to compute the cumulative probability that at least one of the total events corresponds to the signal, and not to the background, as a function of the observation time. This can be done by assuming that the ocurrence of the signal event and backgroungd events $N_{\rm s}$ and $N_{\rm b}$ follow Poisson distributions. As a result, the probability that out of a total of events $N_{\rm tot}=N_{\rm s}+N_{\rm b}$ we have at most $N_{\rm b}<N_{\rm tot}$ can be expressed as
\beq
P( N_{\rm b} < N_{\rm tot} )= \sum_{k=0}^{N_{\rm tot}-1} \frac{N_{\rm b}^k  \exp{\left(-N_{\rm b}   \right) }}{k!}.
\eeq
We show this cumulative probability in Fig.\ref{FigCDF} for both $\Delta \theta_{\rm bin}=\{0.5^\circ,1^\circ\}$, where it can be seen that if PKS 2155-304 actually emits neutrinos at the level suggested here, a neutrino telescope like the planned KM3NeT would be able to ascribe at least one event to this blazar with a probability very close to one after an observation period of four years. We point out here that a resolution of $0.5^\circ$ is actually conservative (e.g. Kappes et al. 2007, Bersani 2012).

\section{Discussion}
We have implemented a two-{component} model to study the emission of the blazar PKS 2155-304. The quiescent state of electromagnetic emission was associated with the contribution produced in a heavy barionic jet, in which both electrons and protons can be accelerated. Specifically, this low state of emission is dominated by synchrotron and inverse Compton interactions of the primary electrons, and also by $pp$ interactions in the jet. The time-dependent contribution {is produced by multiple shocks in} an internal electron-positron beam, surrounded by the jet. We showed that if shocks are injected in this beam, as would be expected as a consequence of Kelvin-Hemlholtz instabilities, then a variable emission can be generated via SSC interactions, giving rise to gamma-ray light curves similar to those observed by HESS.

 Other important output predicted by the present model is in the form of VHE neutrinos produced by $pp$ interactions. If this accompanying flux remains constantly produced over a period of four years, the detection of at least one event from PKS 2155-304 should be guaranteed in a detector like KM3NeT with a probability very close to one. This would be a very important piece of evidence for hadronic acceleration in the source. Conversly, if the neutrinos are not detected at this level, this would be evidence of a leptonic-dominated source.

\begin{acknowledgements}
 The authors would like to thank Pierre Brun for fruitful discussions on EBL absorption.
 This work had the support from CONICET (PIP 112-200801-00587, and PIP 112-
200901-00078).
\end{acknowledgements}

\end{document}